\documentclass[a4paper]{jpconf}
\usepackage{graphicx}
\begin{document}
\title{Dark matter with photons}

\author{\'Alvaro de la Cruz-Dombriz$^{\,1, 2}$ and Viviana Gammaldi$^{\,3}$}

\address{$^{1}$ Astrophysics, Cosmology and Gravity Centre (ACGC), University of Cape Town, Rondebosch, 7701, South Africa.}
\address{$^{2}$ Department of Mathematics and Applied Mathematics, University of Cape Town, Rondebosch 7701, South Africa.}
\address{$^{3}$ Departamento de F\'{\i}sica Te\'orica I, Universidad Complutense de Madrid, av. Complutense s/n, E-28040 Madrid, Spain.}

\ead{alvaro.delacruz-dombriz@uct.ac.za, vivigamm@pas.ucm.es}

\begin{abstract}
If the present dark matter in the Universe annihilates into
Standard Model particles, it must contribute to the 
gamma ray fluxes detected
on the Earth.
The magnitude of such contribution
depends on the particular dark matter candidate, but certain
features of the produced
spectra
may be analyzed in a rather model-independent
fashion. In this communication 
we briefly revise 
the complete photon spectra coming from WIMP
annihilation into Standard Model particle-antiparticle pairs obtained
by extensive Monte Carlo simulations and consequent
fitting functions presented 
by Dombriz et {\it al.} in a wide range of WIMP masses. In order to illustrate the usefulness 
of these fitting functions,
we mention how these results may be applied to the so-called brane-world theories 
whose fluctuations, the branons, behave as 
WIMPs and therefore may spontaneously annihilate in SM particles. The subsequent 
$\gamma$-rays signal in the framework of dark matter indirect searches from Milky Way dSphs and Galactic Center
may provide first evidences for this scenario. 
\end{abstract}

\section{Introduction}
According to present observations of large scale structures, CMB anisotropies 
and light nuclei abundances, 
dark matter (DM) cannot
be accommodated within the Standard model (SM) of elementary particles.
Indeed, DM presence is a required component on cosmological 
scales, but also to provide a 
satisfactory description of rotational speeds of galaxies,
orbital velocities of galaxies in clusters, gravitational lensing
of background objects by galaxy clusters 
and the temperature distribution of hot gas in galaxies and
clusters of galaxies.
The experimental determination of the DM nature will require the
interplay of collider experiments 
and astrophysical
observations. These searches use to be classified in direct or
indirect searches (see \cite{1} and references in Introduction of \cite{Dombriz_wimps_PRD}). 
Concerning direct ones, the 
elastic scattering of DM particles from nuclei should lead directly 
to observable nuclear recoil signatures although
the weak interactions between DM and the 
standard matter makes DM direct detection 
difficult.

On the other hand, DM might be detected indirectly, by observing
their annihilation products into SM particles. Thus, even if WIMPs
(Weakly Interacting Massive Particles) are stable, two of them may annihilate
into ordinary matter such as quarks
, leptons and gauge bosons. Their annihilation in different places (galactic halo, Sun, etc.)
produce cosmic rays to be discriminated through distinctive
signatures from the background. After WIMPs annihilation a
cascade process occurs. In the end the 
stable particles: neutrinos, gamma rays, antimatter... 
may be observed through different devices. Neutrinos 
and gamma rays have the advantage of maintaining their original
direction due to their null electric charges. 
%

This communication precisely focuses 
on photon production coming 
from WIMPs when they annihilate into SM particles.
Photon fluxes in specific DM models are usually obtained
by software packages such as DarkSUSY and micrOMEGAs 
based on PYTHIA Monte Carlo event generator \cite{PYTHIA} 
after having fixed a WIMP mass 
for the particular SUSY model under consideration. 
In this sense, the aim of this investigation
is to provide
fitting functions for the photon
spectra corresponding to each individual SM annihilation channel
and, in addition, determine the dependence of such spectra on the
WIMP mass in  a model independent way. This would allow to
apply the results to alternative DM candidates for which
software packages have not been developed.
On the other hand, the information about channel contribution and
mass dependence can be very useful in order to identify gamma-ray
signals for specific WIMP candidates and may also provide
relevant information about the photon energy distribution 
when SM pairs annihilate.

The paper is organized as follows: in  section 2, we review
the standard procedure for the calculation of gamma-ray fluxes coming from
WIMP pair annihilations. 
Section 3 is then devoted to
the details of spectra simulations performed with PYTHIA. We mention here some important issues about the final state radiation 
and the particular case of the top quark annihilation channel. 
In section 4, we introduce the fitting formulas used to
describe the spectra depending upon the annihilation channel. 
Then, in section 5 we explicitly provide the results for two relevant annihilation channels.
Section 6 is finally devoted to describe how these results may be useful in brane-world theories providing 
WIMP candidates.
\section{Gamma-ray flux from DM annihilation}
Let us remind that the $\gamma$-ray flux from the annihilation 
of two WIMPs of mass $M$ into two SM 
particles coming from all possible annihilation channels (labelled by the subindex $i$) is given by:
\begin{eqnarray}
\label{eq:integrand}
\frac{{\rm d}\,\Phi_{\gamma}^{{\rm DM}}}{{\rm d}\,E_{\gamma}}
 &=& \frac{1}{4\pi M^2}
 \sum_i\langle\sigma_i v\rangle
\frac{{\rm d}\,N_\gamma^i}{{\rm d}\,E_{\gamma}}
 \;\;\;\, \times \;\;\;\,
 \frac{1}{\Delta\Omega} \int_{\Delta\Omega} {\rm d}\Omega
 \int_{\rm l.o.s.} \rho^2 [r(s)]\ {\rm d}s\;,
\\
&&
\underbrace{\;\;\;\;\;\;\;\;\;\;\;\;\;\;\;\;\;\;\;\;\;\;\;\;\;\;\;\;\;\;\;\;\;\;\;}
_{{\rm Particle~model~dependent}}\;\;\;\;\;\;\;\;\;\;
\underbrace{\;\;\;\;\;\;\;\;\;\;\;\;\;\;\;\;\;\;\;\;\;\;\;\;\;\;\;\;\;\;\;\;\;\;\;\;\;\;\;\;\;\;\;}_{{\rm Dark~matter~density~dependent}}\nonumber
\end{eqnarray}
where $M$ denotes the mass of the WIMP,  
$\langle \sigma_{i} v \rangle$ holds for
the thermal averaged annihilation cross-section of two 
WIMPs into two ($i^{th}$ channel) SM 
particles and $\rho$ is the DM density.
%

The first piece of the r.h.s. in (1)
depends upon the particular particle physics model for DM
annihilations. In particular, the self-annihilation cross sections are
mainly described
by the theory explaining the WIMP physics, whereas the number
of photons produced in each decaying channel per energy interval
involves decays and/or hadronization of unstable products,
for instance quarks and gauge bosons. Consequently,
the detailed study of these decay chains
and non-perturbative effects related to QCD is
an almost impossible task to be tackled by any analytical approach.
The second piece in  (\ref{eq:integrand}) is a line-of-sight (l.o.s.)
integration through the DM density distribution of the target and averaged 
over the detector solid angle $\Delta\Omega$. Let 
us discuss each of these pieces separately:

\subsection{Particle Physics model}
Although annihilation cross sections are not known, they are restricted
by collider constraints and direct detection. In addition,
the thermal relic density in the range $\Omega_{\rm CDM}
h^2 = 0.1123\pm 0.0035$ which is determined by fitting the standard
$\Lambda$CDM model to the WMAP7 data 
\cite{Komatsu:2010fb},
the latest measurements from the BAO 
in the distribution of galaxies \cite{Percival:2009} and the Hubble
constant measurement \cite{Riess:2009pu}, do not allow 
arbitrary contributions from the DM gamma ray fluxes.

As already mentioned, the annihilation of WIMPs is closely related
to SM particle production. The time scale of annihilation
processes is shorter than typical astrophysical scales. This fact
implies that only stable or very long-lived particles survive
to the WIMP annihilations and may therefore be observed by detectors.

For most of the DM candidates, the production of mono-energetic photons
is very suppressed.  The main reason for such a suppression comes
from the fact that DM is neutral.
Therefore, the gamma-ray signal comes
fundamentally from secondary
photons originated in the cascade of decays of gauge bosons and jets 
produced from WIMP annihilations. These
annihilations would  produce in the end a broad energy distribution of photons,
which would be difficult to be distinguished from the background.
However, the directional dependence of the gamma ray intensity coming from these
annihilations is mainly localized in point-like sources  
which may 
provide a distinctive signature.
%

All those channels contributions produce a broad energy gamma ray flux, whose
maximum constitutes a potential signature for its detection. 
On the other hand, a different strategy can be followed by taking
into account the fact that the cosmic ray background is suppressed
at high energies. Primary photons coming from the
Weicks\"{a}cker-Williams radiation dominate the spectrum at energies
close to the mass of the DM candidate and their signature is
potentially observable as a cut-off \cite{cutoff}. This approach
is less sensitive to electroweak corrections
which may be important if the mass of the DM candidate is larger
than the electroweak scale \cite{Ciafaloni:2010ti}.

\subsection{DM density directionality}
The line of sight integration can be obtained from:
\begin{equation}
\langle J \rangle_{\Delta \Omega} \doteq \frac{1}{\Delta \Omega}
\int_{\Delta \Omega} J(\psi) {\rm d}\Omega = \frac{2\pi}{\Delta\Omega}
\int_0^{\theta_{\rm max}} {\rm d}\theta\,\sin\theta
\int_{s_{\rm min}}^{s_{\rm max}}
{\rm d}s\,\rho^2 \left(\sqrt{s^2+s_0^2-2 s s_0 \cos\theta}\right)
\label{eq:jav}
\end{equation}
where
$J(\psi) = \int_{\rm l.o.s.} {\rm d}s\,\rho^2(r)$.
The angled brackets denote the averaging over the solid
angle $\Delta \Omega$, and $s_{\rm min}$ and $s_{\rm max}$ are the
lower and upper limits of the line-of-sight integration: $s_0
\cos\theta \pm \sqrt{r_t^2 - s_0^2 \sin^2\theta}$. In this formula
$s_0$ is the heliocentric distance and $r_t$ is the tidal radius.

Traditionally, the galactic center has attracted the attention
of this type of directional analysis since standard cusped
Navarro-Frenk-White halos predict the existence of a very
important amount of DM in that direction 
\cite{stoehr}.
However, this assumption is in contradiction with a substantial body
of astrophysical evidences \cite{evidences}, and a core profile is
not sensitive to standard DM candidates. On the contrary, cusped profiles
are not excluded for the Local Group dwarf spheroidals (dSphs) that constitute
interesting targets since they are much more dominated by DM.  In
this way, directional analysis towards Canis Major, Draco and
Sagittarius or Segue 1 \cite{dSphs} are more promising.

In any case, galaxy clusters are also promising targets \cite{Ref2_referee}. Other alternative strategy 
takes advantage of the large field of view of FERMI,
that may be sensitive to the continuum photon flux coming from DM annihilation
at moderate latitudes ($|b| > 10^\circ$) \cite{stoehr}. Other proposed targets,
as the Large Magellanic Cloud \cite{olinto}, are less interesting since
their central parts are dominated by baryonic matter.

\section{Procedure
}

In this section, we explicitly specify how gamma rays spectra have
been generated and we will discuss some issues concerning the final 
state radiation and the top quark channel particularities.
 
\subsection{Spectra generation}
Throughout this investigation, we have used the particle physics PYTHIA software 
[3] to obtain our results.
%
%
The WIMP annihilation is usually 
split into two separated processes: The first describes the
annihilation of WIMPs and its SM output.
The second one considers the evolution 
of the obtained SM unstable products.
Due to the expected velocity dispersion of DM,
most of the annihilations happen quasi-statically. This fact 
allows to state that by considering different  
center of mass (CM) energies for the obtained SM particles pairs 
from WIMP annihilation process, we are
indeed studying different WIMP masses, i.e. $E_{{\rm CM}} \simeq 2\,M$. 
The procedure to obtain the photon spectra is thus straightforward: For a given pair of 
SM particles which are produced in the WIMP annihilation, we count the produced number of photons. 
Statistics have to be large enough, in particular 
for highly energetic photons usually suppressed when not high enough number of annihilations is simulated.
%


%
%
\subsection{Final State Radiation}
If the final state in the annihilation process contains charged
particles, there is a finite probability of emission of an
additional photon \cite{Bringmann}. In principle there are two 
types of contributions: that coming from photons directly radiated from the 
external legs, which is the final state radiation we have considered in the work, and that coming from virtual
particles exchanged in the WIMP annihilation process. The first kind of
contribution can be described for relativistic final states by means of
an universal Weizs\"{a}cker-Williams term fundamentally independent
from the particle physics model \cite{Bringmann}. On the other hand, radiation from virtual particles only takes place
in certain DM models and is only relevant in
particular cases, for instance, when the virtual particle mass is
almost degenerate with the WIMP mass. Even in these cases,
it has been shown \cite{Cannoni} that although this effect has to be
included for the complete evaluation of fluxes of high energy
photons from WIMP annihilation, its contribution is
relevant only in models and at energies where the lines contribution
is dominant over the secondary photons. For those reasons and
since the aim of the present work
is to provide model independent results for photon spectra,
 only final state radiation was included in our simulations.

\subsection{The case for $t$ quark decay}
The decay of top ($t$) quark is not explicitly included in
PYTHIA package. We have approximated this process by its dominant
SM decay, i.e. each (anti) top decays into $W^{+(-)}$ and (anti) bottom.
In order to maintain any non-perturbative effect, we work on an
initial four-particle state composed by $W^{+} b$ coming
from the top and  $W^{-} \bar{b}$ from anti-top, which keeps all
kinematics and color properties from the original pair.
Starting from this configuration, we have forced decays and
hadronization processes to evolve as PYTHIA does and
therefore, the gamma rays spectra  corresponding to this channel
have also been included in our analysis.

In order to verify the validity of these results, further calculations were 
made \cite{Roberto_2011} by including hadronic string between $b\bar{b}$ 
and improving statistics in the high energy photons range. Conclusions about $t$ 
quark channel remained unchanged with respect to the ones in \cite{Dombriz_wimps_PRD}.


\section{Analytical fits to PYTHIA simulation spectra}
%
In this section we present the  fitting functions used for the different
channels. According to the PYTHIA simulations
described in the previous section, three different parametrizations
were required in order to fit all available data from the studied
channels:  one for quarks (except top quark) and 
leptons, a second one for $W$ and $Z$ gauge bosons and a third one for top quark.

The parameters in the following expressions
were considered in principle to be WIMP mass dependent 
and their  mass dependences 
were fitted by using power laws.



\subsection{Quarks and leptons}
%
%
For quarks (except the top), $\tau$ and $\mu$ leptons,
the most general formula needed to reproduce the
 behavior of the differential
number of photons per photon energy may
be written as:
\begin{eqnarray}
x^{1.5}\frac{{\rm d}N_{\gamma}}{{\rm d}x}\,=\, a_{1}{\rm exp}\left(-b_{1} x^{n_1}-b_2 x^{n_2} -\frac{c_{1}}{x^{d_1}}+\frac{c_2}{x^{d_2}}\right) + q\,x^{1.5}\,{\rm ln}\left[p(1-x)\right]\frac{x^2-2x+2}{x}
\label{general_formula_1}
\end{eqnarray}
In this formula, the logarithmic term
takes into account the final state radiation through the
Weizs\"{a}cker-Williams expression
\cite{Hooper2004,Bringmann}.
Nevertheless, initial radiation
is removed from our Monte Carlo
simulations in order to avoid wrongly counting their
possible contributions.

Strictly speaking, the $p$ parameter in the
Weizs\"{a}cker-Williams term in the previous
formula is $(M/m_{particle})^2$
where $m_{particle}$ is  the mass of the
charged particle that emits radiation. However in our case, it will be
a free parameter to be fitted since the radiation comes
from many possible charged particles, which are produced
along the decay and hadronization processes. Therefore we are
encapsulating all the bremsstrahlung effects in a
single Weizs\"{a}cker-Williams-like term.

Concerning the $\mu$ lepton, the expression above
(\ref{general_formula_1}) becomes simpler
 since the exponential contribution is absent. Thus its flux becomes 
%
%
\begin{eqnarray}
x^{1.5}\frac{{\rm d}N_{\gamma}}{{\rm d}x}\,=\, q\,x^{1.5}\,{\rm ln}\left[p(1-x^{l})\right]\frac{x^2-2x+2}{x}
\label{general_formula_mu}
\end{eqnarray}
where the $l$ parameter in the logarithm is needed
in order to fit the simulations as will be seen in the corresponding
sections.

%
Parameters in expression  (\ref{general_formula_1}) are channel dependent as 
can be found in \cite{Dombriz_wimps_PRD}. Depending on the studied channel, these parameters
may be either dependent or independent from the studied WIMP mass. 
For instance, the case for $c$ quark 
was studied in \cite{Dombriz_QCHSIX}.

\subsection{$W$ and $Z$ bosons}
For the $W$ and $Z$ gauge bosons, the parametrization used to
 fit the Monte Carlo simulation is:
\begin{eqnarray}
x^{1.5}\frac{{\rm d}N_{\gamma}}{{\rm d}x}\,=\, a_{1}\,{\rm exp}\left(-b_{1}\, x^{n_1}-\frac{c_{1}}{x^{d_1}}\right)\left\{\frac{{\rm ln}[p(j-x)]}{{\rm ln}\,p}\right\}^{q}
\label{general_formula_W_Z}
\end{eqnarray}
This expression differs from the expression (\ref{general_formula_1}) in
the absence of the  additive logarithmic contribution. Nonetheless,
this contribution acquires a multiplicative character. The exponential
contribution is also quite simplified with only one positive and one
negative power laws. Moreover, $a_1$, $n_1$ and $q$ parameters
appear to be independent of the WIMP mass.
The rest of parameters, i.e., $b_1$, $c_1$, $d_1$, $p$ and $j$, are WIMP mass dependent and were
determined in \cite{Dombriz_wimps_PRD} for each WIMP mass and for the $W$ and $Z$ separately.
In  both cases the covered WIMP mass range was from
$100$ to $10^4$ ${\rm GeV}$. Nonetheless, for masses higher
than 1000 GeV, no significant
change in the photon spectra for both particles \cite{Dombriz_wimps_PRD} was observed.


\subsection{$t$ quark}
Finally, for the top channel, the required parametrization turned out to be:
\begin{eqnarray}
x^{1.5}\frac{{\rm d}N_{\gamma}}{{\rm d}x}\,=\, a_{1}\,{\rm exp}\left(-b_{1}\, x^{n_1}-\frac{c_{1}}{x^{d_1}}-\frac{c_{2}}{x^{d_2}}\right)\left\{\frac{{\rm ln}[p(1-x^{l})]}{{\rm ln}\,p}\right\}^{q}
\label{general_formula_t}
\end{eqnarray}
%
Likewise the previous case for $W$ and $Z$ bosons,
gamma-ray spectra parametrization for the top is quite
different from that given by expression (\ref{general_formula_1}).
This time, the exponential contribution is more complicated than the
one in expression (\ref{general_formula_W_Z}), with one positive and
two negative power laws. Again, the  additive logarithmic contribution
is absent but it acquires a multiplicative behavior. Notice the
exponent $l$ in the logarithmic argument, which is required to provide correct
fits for this particle.
Moreover, $a_1$, $c_1$,  $d_1$ and $d_2$ parameters
appear to be independent of the WIMP mass.
The rest of parameters, i.e., $b_1$, $n_1$, $c_2$, $p$, $q$ and $j$, are WIMP mass dependent and were
determined in \cite{Dombriz_wimps_PRD}. 
The covered WIMP mass range for the top case was from $200$ 
to $10^5$ GeV. Nevertheless, at masses higher than
1000 GeV it was observed again \cite{Dombriz_wimps_PRD} that there is no significant
change in the  gamma-ray spectra. Consistency of this result was verified in \cite{Roberto_2011}.

\section{Some results: $\tau$ lepton and $t$ quark}
In order to illustrate the explained procedure, we present here 
some representative annihilation channels: $\tau$ lepton and $t$ quark.
 
%
%

For $\tau$ lepton channel the studied mass range was from 25 GeV to 50 TeV. The 
 mass dependent parameters in expression (\ref{general_formula_1}) are
only $n_1$ and $p$ whereas
$a_1$, $b_1$, $b_2$, $n_2$, $c_1$, $d_1$, $c_2$, $d_2$ and $q$ are
mass independent. Figure 1 presents $\tau$ channel spectra for 
1 and 50 TeV WIMP masses.

%

Finally, for $t$ quark channel the studied mass range was from 200 GeV to 10 TeV although the spectra are the same 
from 1 GeV onwards. The mass dependent parameters in expression (\ref{general_formula_t}) are
$b_1$, $n_1$, $c_2$, $p$, $q$ and $l$ whereas
$a_1$, $c_1$, $d_1$ and $d_2$ are
mass independent. Figure 2 presents $t$ channel spectra for two different WIMP masses, 500 and 1000 GeV.


%

\begin{figure}
\includegraphics[height=.235\textheight]{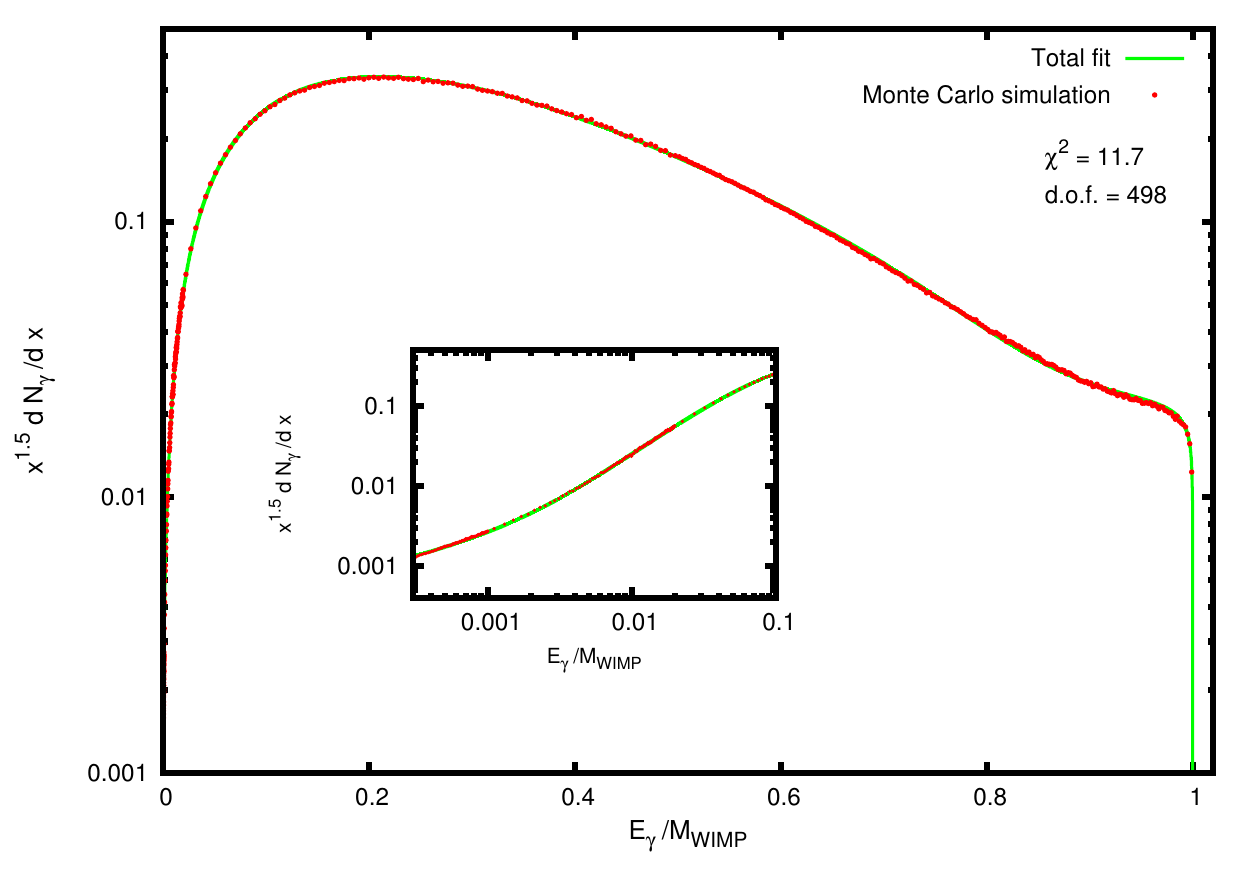}
\includegraphics[height=.235\textheight]{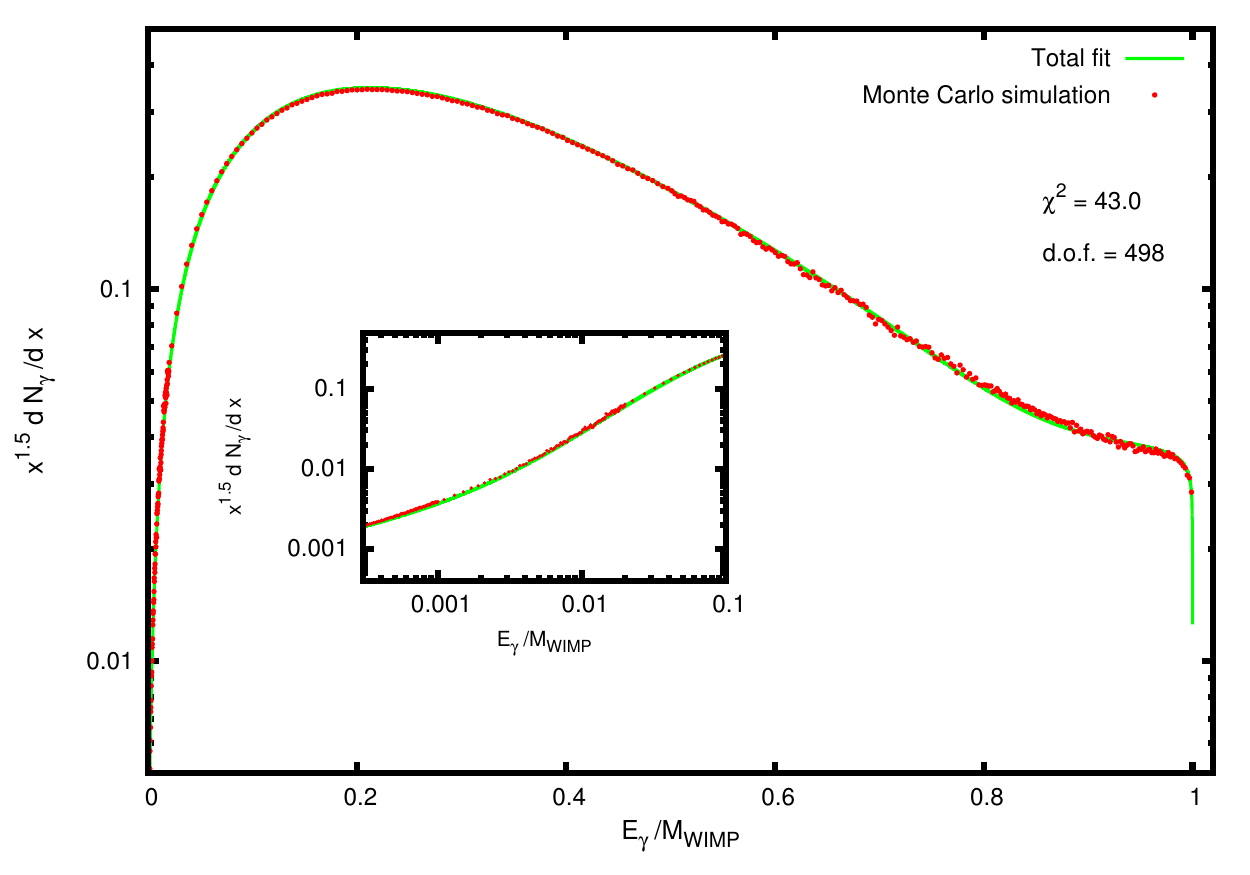}
\caption{Photon spectra for two WIMP masses (1000 GeV and 50 TeV) in the $\tau^{+}\tau^{-}$ 
annihilation channel. Red dotted points are PYTHIA simulations and solid lines correspond to the proposed fitting functions.}
\end{figure}
\begin{figure}
\includegraphics[height=.235\textheight]{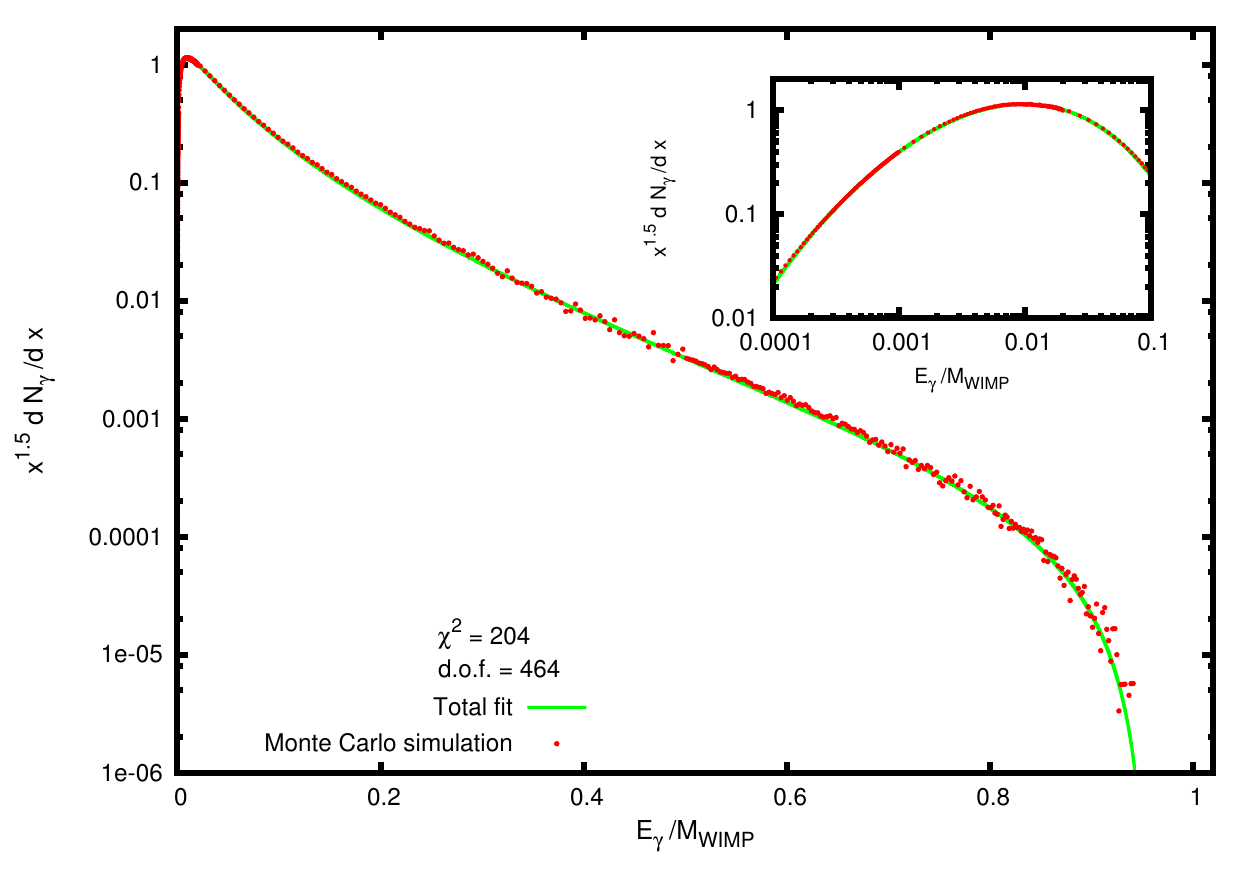}
\includegraphics[height=.235\textheight]{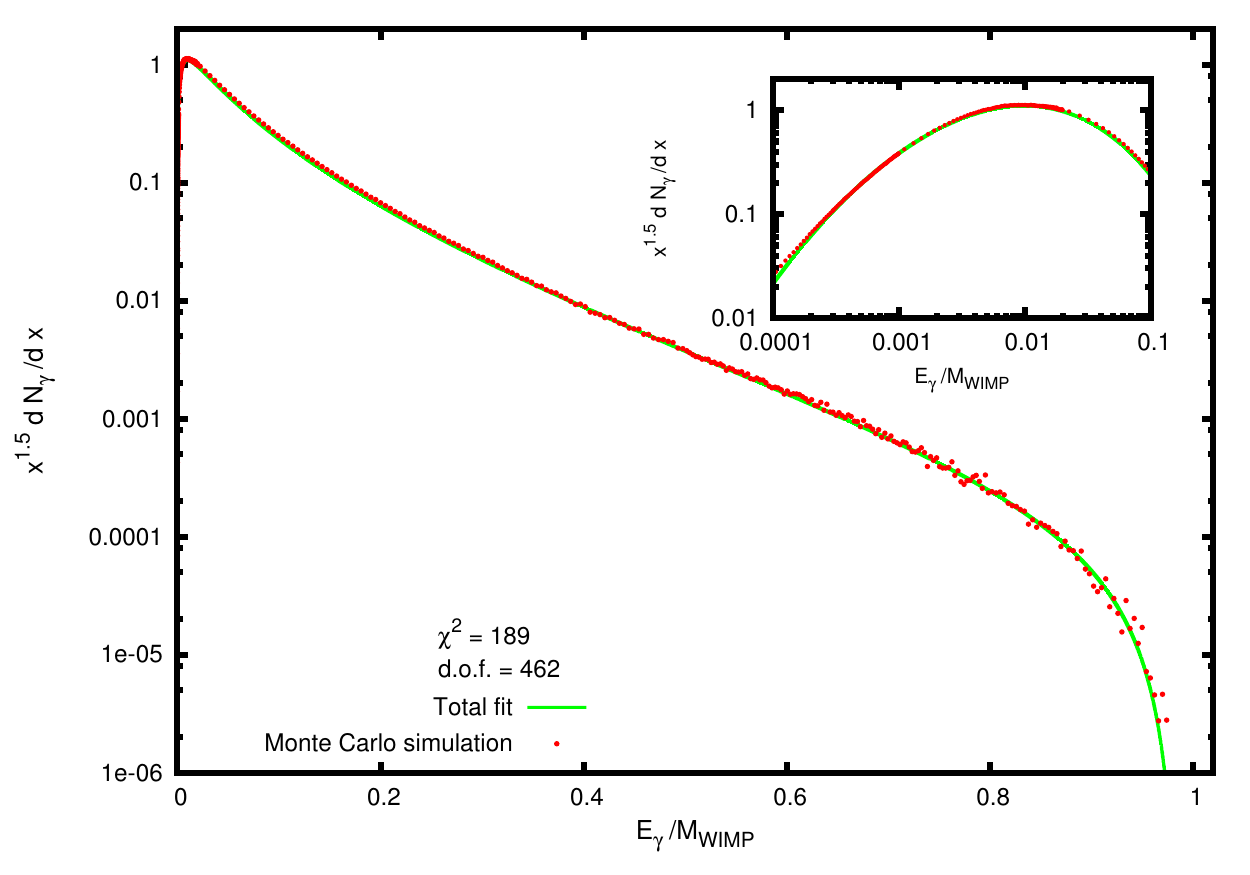}
\caption{Photon spectra for two WIMP masses (500 and 1000 GeV) in the $t\bar{t}$ 
annihilation channel. Red dotted points are PYTHIA simulations and solid lines correspond to the proposed fitting functions.}
\end{figure}
\section{Brane-world theory as an example}

It has been suggested that our universe could be a 3-dimensional brane where the SM fields live embedded 
in a D-dimesional space-time. In flexible braneworlds, in addition to the SM fields, new degrees of freedom appear on the brane associated to brane fluctuations, that is the branons. In brane-world models with low tension, branons appear to be massive and weakly interacting fields, so natural candidates to DM \cite{CDM, BR, ACDM}.  Therefore, their annihilations by pairs may produce SM particles and gamma photons by the subsequent processes of hadronization and decay. Limits on the model 
parameters, the WIMPs mass $M$ and the tension scale $f$, are given both from collider experiments  
and indirect search of DM. 
In particular, the self-annihilation cross section of branons depends on the two parameters of the model \cite{CDM, CDM_cross_sections}. In the case of heavy branons, neglecting three body annihilations and direct production of two photons, the main contribution to the photon flux comes from branon annihilation into 
$ZZ$ and $W^+ W^-$ (see Figure 1 in \cite{gammaflux}) according to the expression
\begin{eqnarray}
\langle\sigma_{Z,W} v\rangle=\frac{M^2\sqrt{\,1-\left(\frac{m_{Z,W}}{M}\right)^2}
\left(4M^4-4M^2m_{Z,W}^2+3m_{Z,W}^4\right)}{64f^8\pi^2}.
\label{Cross_section_WZ}
\end{eqnarray}
The contribution from heavy fermions, i.e. annihilation in $t\bar{t}$ 
channel, can be shown to be subdominant \cite{indirect}. Therefore, expression (\ref{Cross_section_WZ}) represents the 
self-annihilation cross section to be considered in (\ref{eq:integrand}) for the study of these theories.


The astrophysical part of (\ref{eq:integrand}) depends as already mentioned on both the performed experiment and DM profile of the source. $<J>_{\Delta\Omega}$
value is approximately 
$10^{23}\,{\rm GeV^2 cm^{-5} sr^{-1}}$ for dSphs galaxy, but strongly dependent also from the distance of the source for a given DM profile. The technical details of the different experiments and the value of the background also affect the minimum detectable gamma ray flux. The minimum expected value of this flux as 
coming from a given source and instrument may be given by the following expression
\begin{equation}
\frac{\Phi_\gamma\sqrt{\Delta\Omega A_{eff}t}}{\sqrt{\Phi_\gamma+\Phi_{Bg}}} \geq 5, 
\label{minflu}
\end{equation}
%
By integrating expression (\ref{eq:integrand}) over the energy threshold of the selected device, an estimation of 
$N_\gamma<\sigma v>$ can be found and matched with the expected one \cite{gammaflux} depending on the theoretical model. 
This procedure allows to select the most promising target to be investigated with current ground-based or satellites experiments 
(MAGIC \cite{Mag11}, EGRET \cite {EGRET}, FERMI \cite{Fer}) or with a new generation of them (CTAs \cite{CTA}). Refer to   
\cite{gammaflux} for further details.

\section{Conclusions}
We have presented the model-independent fitting functions for the photon spectra coming from
WIMPs pair annihilation into Standard Model particle-antiparticle
pairs for all the phenomenologically relevant channels. 
%
%
%
This analysis is model independent and therefore,
provided a theoretical model 
our formulas make it possible to obtain the expected
photon spectrum 
in a relatively simple way. 
Explicit calculations for all studied
channels [2] are available at the websites \cite{Mathematica_code} and \cite{Fortran_code}.
%
%

\subsection{Acknowledgments}
AdlCD acknowledges financial support from National Research Foundation 
(NRF, South Africa),  MICINN (Spain) project
numbers FIS 2008-01323, FPA 2008-00592 and MICINN
Consolider-Ingenio MULTIDARK CSD2009-00064. AdlCD is particularly grateful to 
PHOTON11 organizing committee for inviting him to present these results.
VG acknowledges financial support from MICINN (Spain)
Consolider-Ingenio MULTIDARK CSD2009-00064.

\end{document}